\documentclass{article}
\usepackage{amssymb}

\usepackage{graphicx}
\usepackage{amsmath}

\input{tcilatex}

\begin{document}

\title{Two photon decay of the pseudoscalars, the chiral symmetry breaking
corrections}
\author{N.F. Nasrallah \\
Faculty of Science, Lebanese University\\
Tripoli, Lebanon}
\date{}
\maketitle

\begin{abstract}
The extrapolation of the decay amplitudes of the pseudoscalar mesons into
two photons from the soft meson limit where it is obtained from the
axial-anomaly to the mass shell involves the contribution of the 0$^{-}$
continuum. These chiral symmetry breaking corrections turn out to be large.
The effects of these corrections on the calculated $\pi ^{0}$ decay rate, on
the values of the singlet-octet mixing angle and on the ratios $\frac{f_{8}}{%
f_{\pi }}$ and$\frac{f_{0}}{f_{\pi }}$ are discussed. The implications for
the transition form factors $\gamma \gamma ^{\ast }\longrightarrow \pi ^{0},$
$\eta ,$ $\eta ^{\prime }$ are also evaluated and confronted with the
available experimental data.
\end{abstract}

\section{Introduction}

Currrent Algebra and PCAC (partial conservation of the axial-vector current)
together with the ABJ axial-anomaly \cite{adler} expresses the two- photon
decay rate of the pseudoscalar mesons ($P=\pi ^{0},\eta ,\eta ^{\prime }$)
in terms of the coupling constants f$_{\pi ^{0}}$, f$_{0}$ and f$_{8}$ and
of the singlet-octet mixing angle $\theta $%
\begin{equation*}
\langle 0\mid A_{\alpha }\mid \pi ^{0}(p)\rangle =-2if_{\pi ^{0}}p_{\alpha }
\end{equation*}
\begin{equation*}
\langle 0\mid A_{\alpha }^{(8)}\mid \eta (p)\rangle =2if_{8}.\cos \theta
.p_{\alpha }\;\;\;,\;\;\;\langle 0\mid A_{\alpha }^{(0)}\mid \eta (p)\rangle
=-2if_{0}.\sin \theta .p_{\alpha }
\end{equation*}
\begin{equation}
\langle 0\mid A_{\alpha }^{(8)}\mid \eta ^{\prime }(p)\rangle =2if_{8}.\sin
\theta .p_{\alpha }\;\;\;,\;\;\;\langle 0\mid A_{\alpha }^{(0)}\mid \eta
^{\prime }(p)\rangle =2if_{0}.\cos \theta .p_{\alpha }  \tag{1.1}
\end{equation}

with 
\begin{equation*}
f_{\pi ^{0}}\simeq f_{\pi ^{+}}=92.4\text{\textrm{MeV}}
\end{equation*}
and where the axial-vector currents are given in terms of the quark fields. 
\begin{equation*}
A_{\alpha }=(\overline{u}\gamma _{\alpha }\gamma _{5}u-\overline{d}\gamma
_{\alpha }\gamma _{5}d)\;\;\;,\;\;\;A_{\alpha }^{(8)}=\frac{1}{\sqrt{3}}(%
\overline{u}\gamma _{\alpha }\gamma _{5}u+\overline{d}\gamma _{\alpha
}\gamma _{5}d-2\overline{s}\gamma _{\alpha }\gamma _{5}s)
\end{equation*}
\begin{equation}
A_{\alpha }^{(0)}=\sqrt{\frac{2}{3}}(\overline{u}\gamma _{\alpha }\gamma
_{5}u+\overline{d}\gamma _{\alpha }\gamma _{5}d+\overline{s}\gamma _{\alpha
}\gamma _{5}s)  \tag{1.2}
\end{equation}
Unlike $f_{\pi },f_{0}$ and $f_{8}$are not directly related to any physical
process.

SU(3) breaking enters through singlet-octet mixing and through the deviation
of the ratio $\frac{f_{8}}{f_{\pi }}$ from unity.Another symmetry breaking
effect originates from SU(3)$\times $SU(3) breaking: the decay rate is
obtained in the soft meson limit from the ABJ anomaly \cite{adler} and the
extrapolation to the mass-shell involves corrections $O(m_{P}^{2})$ which
are expected to be small for the $\pi ^{0}$ but which are not necessarily so
for the $\eta $ and $\eta ^{\prime }$\ These corrections to the PCAC limit,
which arise from the 0$^{-}$ continuum are similar to the corrections to the
Goldberger-Treiman relation \cite{goldberger}. Attempts to estimate these
corrections have been undertaken \cite{kitizawa}, \cite{riazuddin}, \cite
{pham}. In \cite{kitizawa} and \cite{riazuddin} only the contribution of the
high energy part of the $O^{-}$ spectrum was considered and in \cite{pham}
only the low energy part was used. These three calculations are moreover
heavily model dependant.

It is the purpose of the present work to provide an estimate of the PCAC
corrections which enter in the evaluation of the two photon decay rates of
the pseudoscalars under the sole assumptions that the V-A-A vertex is given
by the quark triangle graph in the deep euclidean region and that the main
contribution to SU(3)$\times $SU(3) breaking arises from the energy interval 
\begin{equation*}
1\text{\textrm{GeV}}^{2}\leq s\leq 2\text{\textrm{GeV}}^{2}
\end{equation*}
of the $O^{-}$ continuum.

In section $2$ we give details of the calculation and the resulting
constraints on the values of $f_{8}$, $f_{0}$ and the mixing angle $\theta $
are discussed and finally an evaluation of the implications of our results
on the transition form factors $\gamma ^{\ast }\gamma \rightarrow P$ are
presented in section $3$ and confronted with the available experimental data.

\section{The chiral symmetry breaking corrections}

Consider the three-point function 
\begin{equation*}
T_{\alpha \mu \nu }(p,q_{1},q_{2})=2\pi i.\int \int dxdy\exp
(-iq_{1}x+ipy)\langle 0\mid TA_{\alpha }(y)V_{\mu }(x)V_{\nu }(0)\mid
0\rangle
\end{equation*}
\begin{equation}
=\frac{1}{\pi }.\frac{1}{(p^{2}-m_{\pi }^{2})}.\epsilon _{\mu \nu \lambda
\sigma }q_{1}^{\lambda }q_{2}^{\sigma }p_{\alpha
}.F(p^{2},q_{1}^{2},q_{2}^{2})+...  \tag{2.1}  \label{2.1}
\end{equation}
where the dots represent other tensor structures and where $V_{\mu ,\nu }$%
denotes the electromagnetic current

\begin{equation}
V_{\mu }=\frac{2}{3}\overline{u}\gamma _{\mu }u-\frac{1}{3}\overline{d}%
\gamma _{\mu }d-\frac{1}{3}\overline{s}\gamma _{\mu }s  \tag{2.2}
\end{equation}
The pion pole contribution having been isolated in expression (\ref{2.1}).

In the soft $\pi $ limit and with both photons on mass-shell, the
axial-anomaly \cite{adler} yields 
\begin{equation}
F(0,0,0)=1  \tag{2.3}
\end{equation}
The rate of the decay $\pi ^{0}\rightarrow 2\gamma $ provides a measurement
of 
\begin{equation}
F(m_{\pi }^{2},0,0)=1+\Delta _{\pi },\Delta _{\pi }=O(m_{\pi }^{2}) 
\tag{2.4}
\end{equation}
Equation (\ref{2.1}) is used to define the off-shell symmetric amplitude 
\begin{equation}
F(s,t)=F(s=p^{2},t=q_{1}^{2}=q_{2}^{2})  \tag{2.5}
\end{equation}

In the deep euclidean region $F(s,t)=F^{\text{\textrm{QCD}}}(s,t)$, $F^{%
\text{\textrm{QCD}}}(s,t)$ includes perturbative and non-perturbative
contributions 
\begin{equation}
F^{\text{\textrm{QCD}}}(s,t)=F_{p}(s,t)+F_{np}(s,t)  \tag{2.6}  \label{8}
\end{equation}

The perturbative part $F_{p}(s,t)$ is obtained from the quark triangle
graph, the contribution of which takes a particularly simple form in the
symmetric case \cite{radyushkin} 
\begin{equation}
F_{p}(s,t)=-2.(s-m_{\pi }^{2}).\iint_{0}^{1}dxdy.\frac{x\overline{x}y^{2}}{%
(y(x\overline{x}s-t)+t)},\;\overline{x}=1-x  \tag{2.7}
\end{equation}
The non-perturbative part $F_{np}(s,t)$ arises from the contribution of the
vacuum condensates 
\begin{equation}
F_{np}(s,t)=\frac{b_{1}}{t^{2}}+\frac{b_{2}}{t^{3}}+\frac{b_{3}}{st}+... 
\tag{2.8}  \label{2.8}
\end{equation}
where 
\begin{equation}
b_{1}=-\frac{\pi ^{3}}{9}.\langle \frac{\alpha _{s.}G^{2}}{\pi }\rangle
\;\;\;,\;\;\;b_{2}=-\frac{64}{27}.\pi ^{4}.\langle \alpha
_{s}(qq)^{2}\rangle \;\;\;,\;\;\;b_{3}=-b_{1}  \tag{2.9}
\end{equation}
For t fixed,$F(s,t)$ is an analytic function of the complex variable s with
a cut on the positive real axis which extends from $s=9m_{\pi }^{2}$ to $%
\infty $ \cite{horejsi}$.$

A dispersion relation between $F(s=m_{\pi }^{2},t)$ and $F(0,t)$ is obtained
from the integral 
\begin{equation}
\frac{1}{2\pi i}\int_{c}\frac{ds}{s.(s-m_{\pi }^{2})}.F(s,t)  \tag{2.10}
\label{2.10}
\end{equation}
where $c$ is the closed contour in the complex plane consisting of a circle
of large radius $R$ and two straight lines lying immediatly above and
immediatly below the cut .Cauchy's theorem yields then 
\begin{equation}
F(m_{\pi }^{2},t)=F(0,t)+\frac{m_{\pi }^{2}}{2\pi i}.\int_{c}\frac{ds}{%
s.(s-m_{\pi }^{2})}.F(s,t)  \tag{2.11}
\end{equation}

The integral in the equation above consists of two parts : an integral of
the discontinuity of $F(s,t)$ over the cut , which provides the main
contribution and an integral over the circle of radius $R$ where $F^{\text{%
\textrm{QCD}}}(s,t)$ provides a good approximation to $F(s,t)$ except
possibly in the vicinity of the positive real axis. Little is known about
the integrand over the cut which consists of the contribution of the
axial-vector $(1^{+})$ and pseudoscalar $(0^{-})$ intermediate states.The
contribution of the low energy region(the three pion states) is suppressed
by loop factors and amounts to little \cite{goity-Lewis}. We expect the
major part of the contribution to originate from the $a1(1260)$ and $\pi
^{\prime }(1300)$ bumps i.e from the range $1$\textrm{GeV}$^{2}\leq s\leq $ $%
2$\textrm{GeV}$^{2}.$In order to eliminate this contribution,we shall
consider, instead of integral (\ref{2.10}), the following modified integral 
\begin{equation}
\frac{1}{2\pi i}.\int_{c}\frac{ds}{(s-m_{\pi }^{2})}.(\frac{1}{s}%
-a_{0}-a_{1}.s).F(s,t)  \tag{2.12}
\end{equation}

In the equation above . the coefficients $a_{0}$ and $a_{1}$ will be chosen
so as to annihilate the integrand at $m_{1}^{2}=m_{a1}^{2}=1.56$\textrm{GeV}$%
^{2}$ and at $m_{2}^{2}=m_{\pi ^{\prime }}^{2}=1.70$\textrm{GeV}$^{2}$, i.e 
\begin{equation}
a_{0}=\frac{1}{m_{1}^{2}}+\frac{1}{m_{2}^{2}}\;\;\;,\;\;\;a_{1}=-\frac{1}{%
m_{1}^{2}.m_{2}^{2}}  \tag{2.13}
\end{equation}

This choice reduces the integrand to only a few percent of it's initial
value over the interval $1$\textrm{GeV}$^{2}\leq s\leq $ $2$\textrm{GeV}$%
^{2} $. It will also considerably reduce the contribution to the integral
over the circle near the positive real axis.The contribution of the
continuum having thus been drastically reduced, we neglect it. Cauchy's
theorem now yields 
\begin{equation}
F(m_{\pi }^{2},t)\approx F(0,t)+a_{0}.m_{\pi }^{2}.F(m_{\pi
}^{2},t)+a_{1}.m_{\pi }^{4}.F(m_{\pi }^{2},t)  \notag  \label{16}
\end{equation}
\begin{equation}
+\frac{1}{2\pi i}\oint \frac{ds}{(s-m_{\pi }^{2})}.(\frac{1}{s}%
-a_{0}-a_{1}.s).F(s,t)  \tag{2.14}  \label{2.14}
\end{equation}
The integral being now carried over the circle of radius $R.$

In order to obtain $F(s,0)$ we proceed in a similar fashion.$F(s,t)$ is an
analytic function of the complex variable $t$ except for a cut on the
positive real axis extending from $t=4m_{\pi }^{2}$ to $\infty $.In the low $%
t$ region $F(s,t)$ is dominated by the $\rho -\omega $ double and single
poles, 
\begin{equation}
F(s,t)=\frac{c_{1}(s)}{(t-m_{\rho }^{2})^{2}}+\frac{c_{2}(s)}{(t-m_{\rho
}^{2})}+...  \tag{2.15}
\end{equation}
Consider now the integral 
\begin{equation}
\frac{1}{2\pi i}\int_{c^{\prime }}dt.\frac{(t-m_{\rho }^{2})^{2}}{t}.F(s,t) 
\tag{2.16}  \label{2.16}
\end{equation}
where $c^{\prime }$ is a contour similar to $c$ in the complex $t$ plane.
the double and single vector meson poles having been removed and for an
appropriate choice of $R^{\prime }$ the major contribution to the integral (%
\ref{2.16}) comes from the integral over the circle so that 
\begin{equation}
F(s,0)\thickapprox \frac{1}{m_{\rho }^{4}}.\frac{1}{2\pi i}\oint_{R^{\prime
}}dt\frac{(t-m_{\rho }^{2})^{2}}{t}.F(s,t)  \tag{2.17}  \label{2.17}
\end{equation}
It follows from equations (\ref{2.14}) and (\ref{2.17}) that 
\begin{equation*}
F(m_{\pi }^{2},0).(1-a_{0}.m_{\pi }^{2}-a_{1}.m_{\pi }^{4})
\end{equation*}
\begin{equation}
=F(0,0)+\frac{m_{\pi }^{2}}{(2\pi i)^{2}}\oint \oint \frac{ds\,dt}{%
t(s-m_{\pi }^{2})}(t-m_{p}^{2})^{2}(\frac{1}{s}-a_{0}-a_{1}.s).F^{\text{%
\textrm{QCD}}}(s,t)  \tag{2.18}  \label{2.18}
\end{equation}
The integral above being carried over the circles of large radii $R$ and $%
R^{\prime }$\ we have replaced $F(s,t)$ by $F^{\text{\textrm{QCD}}}(s,t)$ in
the integrand. Inserting eqs (\ref{8}) - (\ref{2.8}) in eq. (\ref{2.18})
gives 
\begin{equation}
F(m_{\pi }^{2},0).(1-a_{0}.m_{\pi }^{2}-a_{1}.m_{\pi }^{4})  \notag
\label{21}
\end{equation}
\begin{equation*}
=F(0,0)+\frac{m_{\pi }^{2}}{9.m_{\rho }^{2}}+\frac{m_{\pi }^{2}}{m_{\rho
}^{4}}.(\frac{R}{5}-\frac{1}{3}(R^{\prime }-2.m_{\rho }^{2}.\ln \frac{%
R^{\prime }}{R}-\frac{m_{\rho }^{4}}{R^{\prime }}))
\end{equation*}
\begin{equation*}
-a_{0}.\frac{m_{\pi }^{2}}{m_{\rho }^{4}}.R.(\frac{R}{10}-\frac{2}{3}%
.m_{\rho }^{2})-a_{1}.\frac{m_{\pi }^{2}.R^{2}}{m_{\rho }^{4}}.(\frac{R}{15}-%
\frac{m_{\rho }^{2}}{3})-\frac{m_{\pi }^{2}}{m_{\rho }^{4}}%
.(a_{0}+a_{1}.m_{\pi }^{2}).b_{1}
\end{equation*}
\begin{equation}
+2.\frac{m_{\pi }^{2}}{m_{\rho }^{2}}.a_{1}.b_{3}  \tag{2.19}  \label{2.19}
\end{equation}
The non-perturbative contributions to equation (\ref{2.19}) turn out to be
negligible.

The decays $\eta \longrightarrow 2\gamma $ and $\eta ^{\prime
}\longrightarrow 2\gamma $ are treated in a similar fashion. The
axial-vector currents which project on the $\eta $ and $\eta ^{\prime }$
states are, respectively, 
\begin{equation}
A_{\alpha }^{\eta }=(\frac{A_{\alpha }^{(8)}}{f_{8}}.\cos \theta -\frac{%
A_{\alpha }^{(0)}}{f_{0}}.\sin \theta )\;\;\;,\;\;\;A_{\alpha }^{\eta
^{\prime }}=(\frac{A_{\alpha }^{(8)}}{f_{8}}.\sin \theta +\frac{A_{\alpha
}^{(0)}}{f_{0}}.\cos \theta )  \tag{2.20}  \label{22}
\end{equation}
These currents we use in the definition of the 3-point function equation (%
\ref{2.1}). The results for the two photon decay rates of the pseudoscalars
are then: 
\begin{equation}
\Gamma (\pi ^{0}\longrightarrow 2\gamma )=\frac{\alpha ^{2}.m_{\pi }^{3}}{%
64\pi ^{3}.f_{\pi ^{+}}^{2}}.(1+\Delta _{\pi })^{2}.\frac{f_{\pi ^{+}}^{2}}{%
f_{\pi ^{0}}^{2}}  \tag{2.21}
\end{equation}
\begin{equation}
\Gamma (\eta \longrightarrow 2\gamma )=\frac{\alpha ^{2}.m_{\eta }^{3}}{%
192\pi ^{3}f_{\pi ^{+}}^{2}}.(1+\Delta _{\eta })^{2}.(\frac{\cos \theta }{%
F_{8}}-2\sqrt{2}.\frac{\sin \theta }{F_{0}})^{2}  \tag{2.22}  \label{2.22}
\end{equation}
\begin{equation}
\Gamma (\eta ^{\prime }\longrightarrow 2\gamma )=\frac{\alpha ^{2}.m_{\eta
^{\prime }}^{3}}{192\pi ^{3}f_{\pi ^{+}}^{2}}.(1+\Delta _{\eta ^{\prime
}})^{2}.(\frac{\sin \theta }{F_{8}}+2\sqrt{2}.\frac{\cos \theta }{F_{0}})^{2}
\tag{2.23}  \label{2.23}
\end{equation}
With the notation $F_{0,8}=$ $\frac{f_{0,8}}{f_{\pi ^{+}}}$ and 
\begin{equation*}
F(m_{\pi }^{2},0,0)=(1+\Delta _{\pi })
\end{equation*}
\begin{equation}
F(m_{\eta }^{2},0,0)=\frac{1}{\sqrt{3}}(1+\Delta _{\eta
})\;\;\;,\;\;\;F(m_{\eta ^{\prime }}^{2},0,0)=\frac{2\sqrt{2}}{3}(1+\Delta
_{\eta ^{\prime }})  \tag{2.24}  \label{2.24}
\end{equation}
$\Delta _{\pi }$ is obtained from equation (\ref{2.19})\ \ The same equation
with the negligible non-perturbative part omitted yields $\Delta _{\eta }$
and $\Delta _{\eta ^{\prime }}$when $m_{\pi }$ is replaced by $m_{\eta ,\eta
^{\prime }}$and when the strange quark mass is neglected.

$R^{\prime },$ the duality radius in the $\rho $-meson channel is usually
taken to be $R^{\prime }\thickapprox $ $1.5$\textrm{GeV}$^{2}$ in the
litterature \cite{shifman} \ The value of $R$ should be large enough to
include the contribution of the pseudoscalar excitations but not too large
in order not to invalidate the approximations at hand.\ Moreover \ $\Delta
_{P}$ should be stable against small variations in $R..$For the parameters $%
m_{1}$ and $m_{2}$ we take 
\begin{equation}
m_{1}^{2}=m_{a_{1}(1260)}^{2}=1.56\text{\textrm{GeV}}^{2}\;\;\;,\;\;%
\;m_{2}^{2}=m_{\pi (1300)}^{2}=1.70\text{\textrm{GeV}}^{2}  \tag{2.25}
\end{equation}
as discussed previously .

In the $\eta $ and $\eta ^{\prime }$ channels two pseudoscalar excitations, $%
\eta (1295)$ and $\eta (1440)$ as well as two axial-vectors, $\ f_{1}(1285)$
and $\ f_{1}(1420)$ dominate the $0^{-}$ and $1^{+}$ continua and couple to
both $\eta $ and $\eta ^{\prime }$ mesons with unknown strengths. Because $%
\eta (1295)$ and $f_{1}(1285)$ on the one hand and $\eta (1440)$ and $%
f_{1}(1420)$ on the other hand are practically degenerate in mass it
suffices to take

\begin{equation}
m_{1}^{2}=1.66\text{\textrm{GeV}}^{2}\;\;\;,\;\;\;m_{2}^{2}=2.04\,\text{%
\textrm{GeV}}^{2}  \tag{2.26}
\end{equation}
for both $\eta $ and $\eta ^{\prime }.$

With these choices expression (\ref{2.19}) passes through a maximum for $%
R\backsimeq $ $2$\textrm{GeV}$^{2}$ in the $\pi $ channel and $R\backsimeq $ 
$2.25$\textrm{GeV}$^{2}$ in the $\eta ,\eta ^{\prime }$ channels.,values
which we adopt because of the stability criteria stated above. Numerically
then, we find 
\begin{equation}
\Delta _{\pi }\backsimeq .047\;\;\;,\;\;\;\Delta _{\eta }\backsimeq
.77\;\;\;,\;\;\;\Delta _{\eta ^{\prime }}\backsimeq 6.0  \tag{2.27}
\label{2.27}
\end{equation}
The relative theoretical error on the numbers above is estimated to be of
the order of $\alpha _{s}(2.5$ \textrm{GeV}$^{2})/\pi $=$12\%$

The chiral symmetry breaking corrections are thus seen to be quite large.For
the $\pi ^{0}$the theoretical value of the width is increased from $\Gamma
_{\pi ^{0}\rightarrow 2\gamma }=7.74$ \textrm{eV} to 
\begin{equation}
\Gamma _{\pi ^{0}\rightarrow 2\gamma }=(8.25\pm .09)\,\text{\textrm{eV}} 
\tag{2.28}
\end{equation}
if we take $f_{\pi ^{+}}/f_{\pi ^{0}}=1.$ Another correction to the chiral
limit could originate from $\pi ^{0},\eta $ and $\eta ^{\prime }$ mixing.It
has been estimated in chiral perturbation theory \cite{Goity-Bernstein} and
would increase the decay rate by another 1.045 factor to 
\begin{equation}
\Gamma _{\pi ^{0}\rightarrow 2\gamma }=(8.60\pm .10)\,\text{\textrm{eV}} 
\tag{2.29}
\end{equation}
The current world average experimental value is 
\begin{equation}
\Gamma _{\pi ^{0}\rightarrow 2\gamma }=(7.75\pm .60)\,\text{\textrm{eV}} 
\tag{2.30}  \label{2.30}
\end{equation}
The large dispersion of the experimental \ results however suggests that the
quoted error is underestimated \cite{Goity-Bernstein}. The outcome of the \
PRIMEX experiment at Jefferson Lab \cite{Gasparian}, which aims at a
precision of 1.5\% in the measurement of the $\pi ^{0}$ decay rate,is thus
eagerly expected in order to clarify the situation.

In order to extract information from our results on the $\eta $ and $\eta
^{\prime }$ more input has to be added

An analysis of the $\eta ,$ $\eta ^{\prime }$mesons mass matrix together
with chiral perturbation theory \cite{donoghue} yields for the mixing angle $%
\theta =-19.5^{\circ }$, this value and the experimentally measured decay
rates, when inserted in equations (\ref{2.22}), (\ref{2.23}) and (\ref{2.27}%
) imply 
\begin{equation}
F_{8}=1.29\pm .09\;\;\;,\;\;\;F_{0}=4.70\pm .45  \tag{2.31}  \label{2.31}
\end{equation}
Our result for $F_{8}$ is consistent with that obtained from chiral
perturbation theory,$F_{8}=1.25$ \cite{donoghue}, the value obtained for$%
F_{0}$ is quite larger than the ones appearing in the litterature.It is
worth noting that a pure gluon component present in $\eta ,\eta ^{\prime }$%
would yield the value appearing in eq.(\ref{2.31}) as an upper limit for $%
F_{0}$ \cite{riazuddin}.

It is unfortunate that $F_{8}$ and $F_{0}$ are not directly linked to any
physically measurable quantity. Data exist however on the transitions $%
\gamma \gamma ^{\ast }\longrightarrow P$ \cite{cello}$.$ We shall examine in
the next section the implications of our results on the corresponding form
factors .

\section{Form factors of the transitions $\protect\gamma \protect\gamma %
^{\ast }\longrightarrow \protect\pi ^{0},\protect\eta $ $,\protect\eta
^{\prime }$}

In the deep euclidean region we have \cite{radyushkin} 
\begin{equation}
F_{p}(s,Q^{2},t)=-2.(s-m_{P}^{2}).\int_{0}^{1}dx\,x\overline{x}\int_{0}^{1}dy%
\frac{y^{2}}{(y.(x\overline{x}s+xQ^{2}+\overline{x}q^{2})-xq^{2}+\overline{x}%
t)}  \tag{3.1}  \label{3.1}
\end{equation}
$t=q_{1}^{2}$, $Q^{2}=-q_{2}^{2}$ and $s=p^{2}$ as before.In order to be
able to compare with experiment, we need to evaluate $F(m_{P}^{2},Q^{2},0).$
The method used in the preceding section,,modified to take into account the\
fact that we have\ now only\ a single pole in the vector meson channel now
yields 
\begin{equation*}
F(m_{P}^{2},Q^{2},0)
\end{equation*}
\begin{equation}
=-\frac{1}{m_{\rho }^{2}}.\frac{1}{(2\pi i)}.\oint \oint ds\,dt\frac{%
(t-m_{\rho }^{2)}}{t}.(\frac{1}{(s-m_{\rho }^{2}}%
-a_{0}-a_{1}.s).F^{QCD}(s,Q^{2},t)  \tag{3.2}  \label{3.2}
\end{equation}
an expression valid for large $Q^{2}.$ Inserting expression (\ref{3.1}) in
the equation above , it takes some straightforward algebra in the complex
plane to evaluate $F(m_{p}^{2},Q^{2},0)$. The deep inelastic limit $%
Q^{2}\longrightarrow \infty $ is readily obtained 
\begin{equation}
Q^{2}.F_{\pi ^{0}}(m_{\pi }^{2},Q^{2},0)\longrightarrow \frac{(\frac{R^{3}}{3%
}-(m_{1}^{2}+m_{2}^{2}).\frac{R^{2}}{2}+m_{1}^{2}.m_{2}^{2}.R)}{%
(m_{1}^{2}-m_{\pi }^{2}).(m_{2}^{2}-m_{\pi }^{2})}  \tag{3.3}  \label{3.3}
\end{equation}
The corresponding expressions for the $\eta $ and $\eta ^{\prime }$ have to
be multiplied by the factors $\frac{1}{\sqrt{3}}(\frac{\cos \theta }{F_{8}}-%
\frac{2\sqrt{2}\sin \theta }{F_{0}})$ and $\frac{1}{\sqrt{3}}(\frac{\sin
\theta }{F_{8}}-\frac{2\sqrt{2}\cos \theta }{F_{0}})$ respectively.
Numerically then 
\begin{equation}
\frac{Q^{2}.F_{\pi ^{0}}(m_{\pi }^{2},Q^{2},0)}{8.\pi ^{2}.f_{\pi }^{2}}%
\longrightarrow .81  \tag{3.4}  \label{3.4}
\end{equation}
\begin{equation}
\frac{Q^{2}.F_{\eta }(m_{\eta }^{2},Q^{2},0)}{8.\pi ^{2}.f_{\pi }^{2}}%
\longrightarrow (1.34.\pm .12).\frac{1}{\sqrt{3}}.(\frac{\cos \theta }{F_{8}}%
-\frac{2\sqrt{2}.\sin \theta }{F_{0}})  \tag{3.5}
\end{equation}
\begin{equation}
\frac{Q^{2}.F_{\eta ^{\prime }}(m_{\eta ^{\prime }}^{2},Q^{2},0)}{8.\pi
^{2}.f_{\pi }^{2}}\longrightarrow (3.72.\pm .54).\frac{1}{\sqrt{3}}.(\frac{%
\sin \theta }{F_{8}}+\frac{2\sqrt{2}.\cos \theta }{F_{0}})  \tag{3.6}
\end{equation}
From the experimental widths and the values of $\theta ,$ $F_{0},$ $F_{8}$
given by equation (\ref{2.31}) we get

\begin{equation}
\frac{Q^{2}.F_{\eta }(m_{\eta }^{2},Q^{2},0)}{8\pi ^{2}f_{\pi }^{2}}%
\longrightarrow .72\pm .07  \tag{3.7}  \label{3.7}
\end{equation}
\begin{equation}
\frac{Q^{2}.F_{\eta ^{\prime }}(m_{\eta ^{\prime }}^{2},Q^{2},0)}{8\pi
^{2}f_{\pi }^{2}}\longrightarrow .98\pm .14  \tag{3.8}  \label{3.8}
\end{equation}
The outputs of equations (\ref{3.4}), (\ref{3.7}) and (\ref{3.8}) together
with the experimental data \cite{cello} are shown on the figures

\section{Conclusion}

The chiral symmetry breaking corretions to the decays $\pi ^{0},\eta ,\eta
^{\prime }\rightarrow 2\gamma $ have been estimated in a model independant
way and found to be quite large.In the case of $\pi ^{0}\rightarrow 2\gamma $
the theoretical value is onsidertably increased and this lends a great
significance to the upcoming PRIMEX experiment \cite{Gasparian}.

In addition the form factors of the transitions $\gamma ^{\ast }\gamma
\rightarrow \pi ^{0},\eta ,\eta ^{\prime }$ were evaluated and shown to
compare favourably with the available experimental data.\pagebreak

{\LARGE Figure Captions: }

\bigskip

Fig1: The form factors of the transitions:

a) $\gamma \gamma ^{\ast }\rightarrow \pi ^{0}$ \ \ \ , \ \ \ b) $\gamma
\gamma ^{\ast }\rightarrow \eta $ \ \ \ , \ \ \ c) $\gamma \gamma ^{\ast
}\rightarrow \eta ^{\prime }$ as obtained from eq. (\ref{3.2}). The vertical
lines represent the data as taken from \cite{cello}

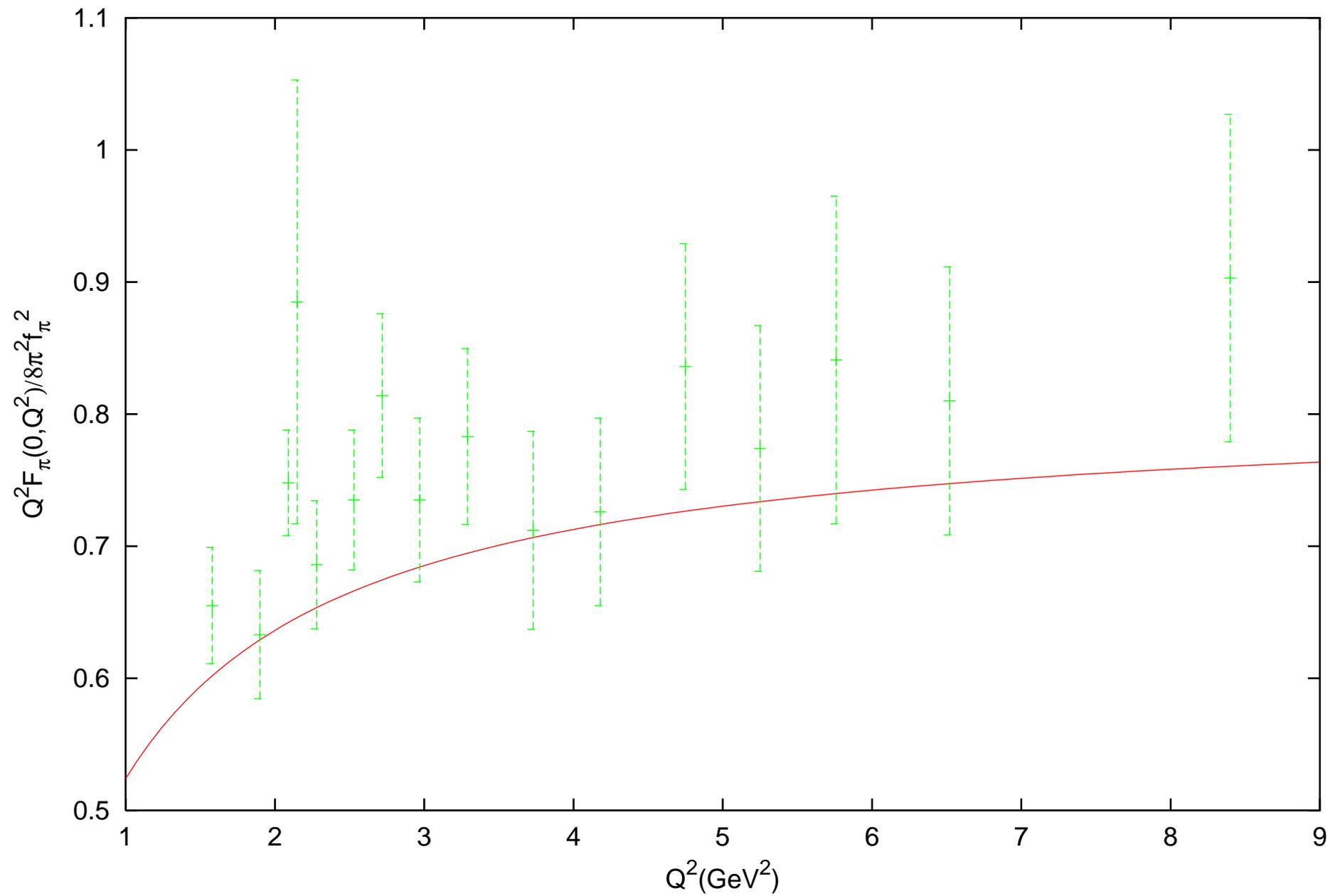
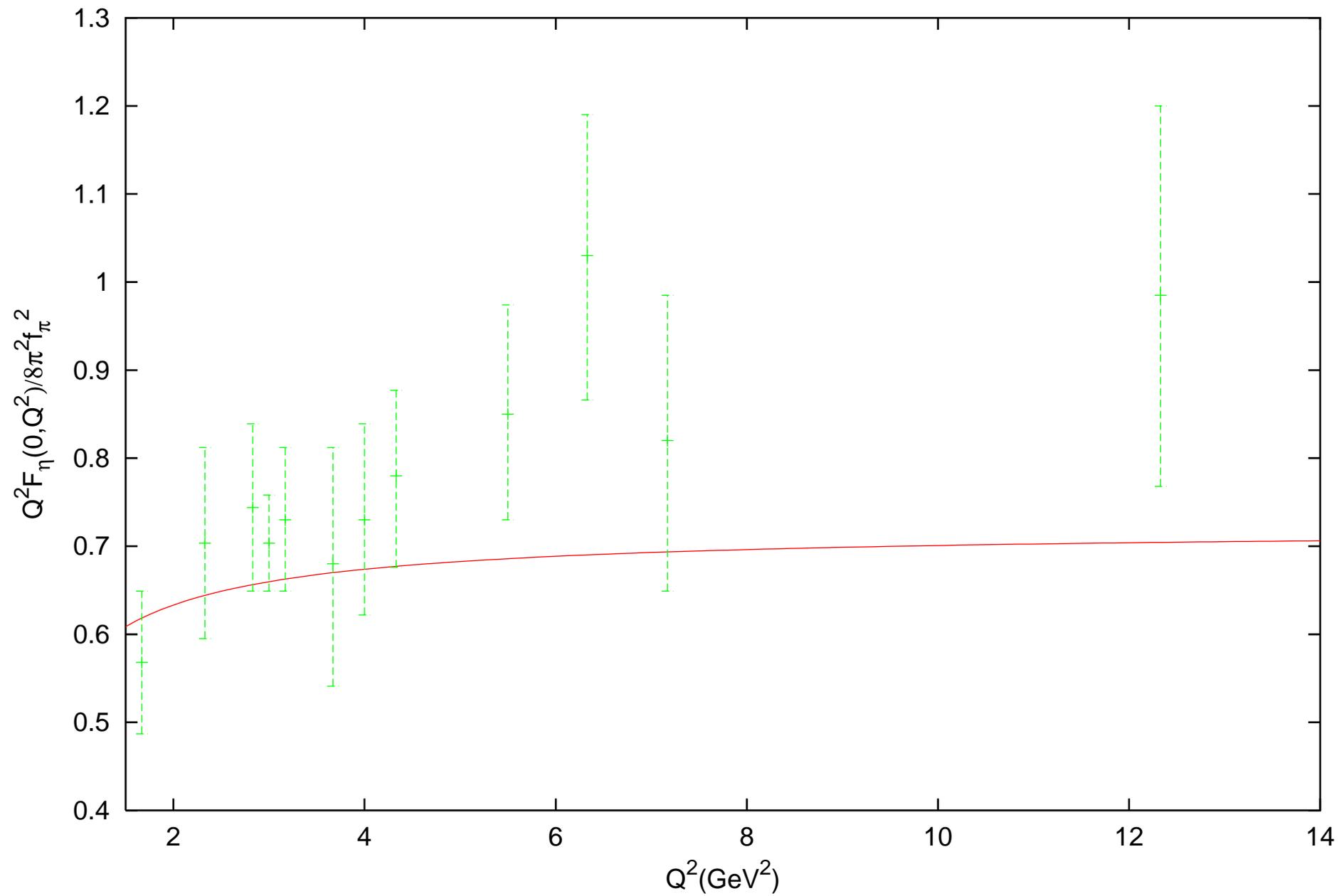
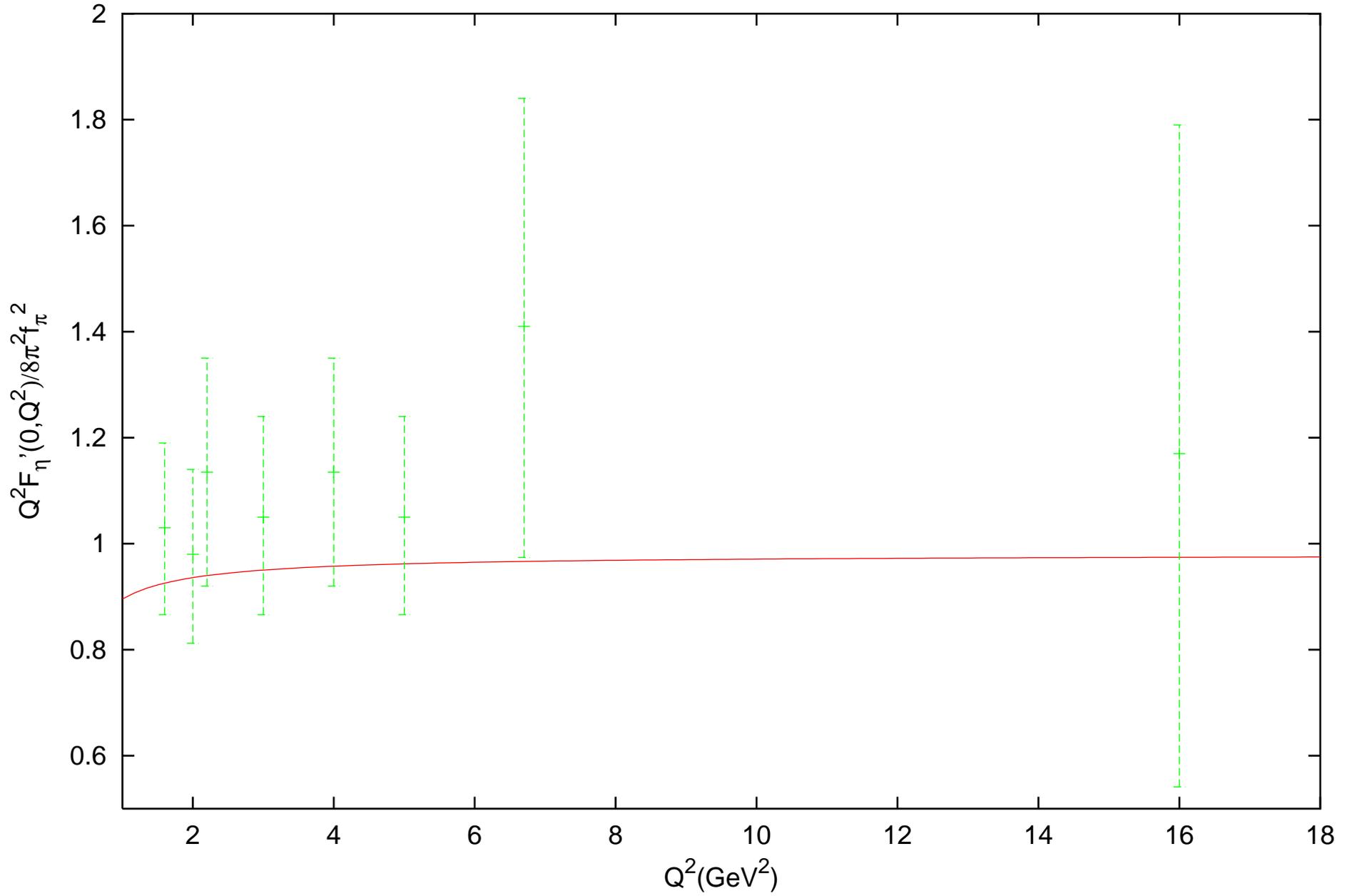
\end{document}